\documentclass[showpacs,aps,prd,nofootinbib,floatfix,amsmath,amssymb]{revtex4}
\usepackage{graphicx}% Include figure files
\usepackage{dcolumn}% Align table columns on decimal point
\usepackage{bm}% bold math
\usepackage{subfigure}
\usepackage{color}
\newcommand{\ebm}{e^{-b\phi}}
\newcommand{\lebm}{\lambda\ebm}

\begin{document}

\title{Sigmoidal Inflation}

\author{E. D\'iaz, O. Meza-Aldama}

\affiliation{Sibatel Communications \\ 303 W Lincoln Ave No.140, Anaheim, CA 92805}%
\date{\today}% It is always \today, today,
             %  but any date may be explicitly specified

\begin{abstract}
In this paper we present a new cosmological inflationary model which is constructed using the Ivanov-Salopek-Bond method with a logistic generating function. We derive the inflationary observables as well as the duration and temperature of the subsequent reheating epoch of our model exactly, with no need to recur to the slow roll approximation. The obtained scalar spectral index and tensor-to-scalar ratio of perturbations fall comfortably within the range of the measurements presented by the Planck collaboration. On the other hand, for the reheating era, our model predicts a relatively small number of e-folds and thus high temperatures, still within range of Planck's bounds.
\end{abstract}

\pacs{98.80.Cq}

\maketitle

\section{Introduction}
The data released by the Planck collaboration \cite{Planck} has high precision cosmological observations which have discarded and restricted inflationary models; of particular relevance is the spectral index of scalar perturbations $n_s$ = $0.9649\pm0.0042$ at 68\% CL and the upper limit on the tensor-to-scalar ratio $r<0.056$. These new measurements allow us to validate new inflationary models, particularly for models which follow slow roll dynamics. However, we find that by using the so called Ivanov-Salopek-Bond (ISB) formalism \cite{chervon} it is not necessary to use the slow-roll approximation in the construction of the model, additionally the work involving the slow-roll parameters is greatly simplified since they are defined purely in terms of the Hubble parameter and its time derivatives. The formalism has been previously employed with generating functions that were of a polynomial form, trigonometric, exponential \cite{Ivanov}, inverse potential and hyperbolic \cite{Muslimov}. While we make use of a logistic generating function for the effective potential during inflation and reheating, this form of the generating function allowed us to construct a potential that complies with current Planck measurements of the tensor to scalar ratio and spectral index. Furthermore, the reheating period of the universe is also considered, since it is important for the subsequent evolution of the universe that the inflaton, through its decays, gives rise to the Standard Model matter content.

\indent This work is organized as follows: in Section 2 we present the results of using a logistic generating function to calculate the effective potential during inflation, and describe its qualitative features. In Section 3 we calculate the spectral index, the tensor-to-scalar ratio, and scalar power spectrum produced during inflation. In Section 4 we obtain the duration of the reheating period in terms of the number of e-folds, as well as the final temperature of this era. Finally, we present our conclusions in Section 5.

\section{A logistic generating function}

\begin{figure}[h!]
	\centering
	\includegraphics[scale=0.35]{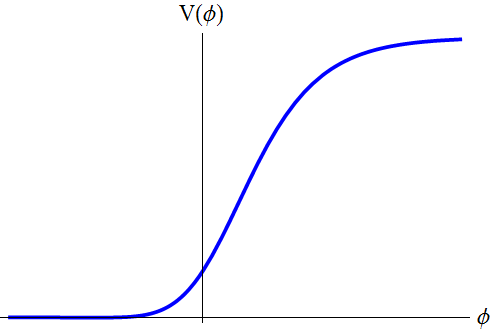}
	\caption{The sigmoidal profile of the potential presented in equation \ref{eq:potential_c0} at $b^2 \leq 3 \kappa / 2$.\label{fig:pot_sig}}
\end{figure}

\begin{figure}[h!]
	\centering
	\includegraphics[scale=0.35]{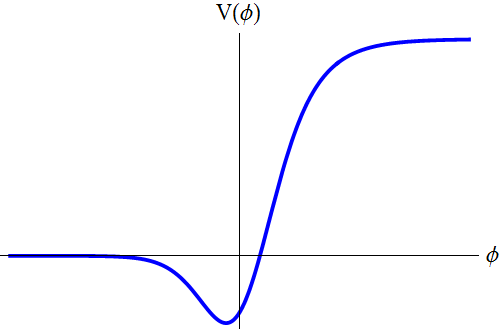}
	\caption{The potential develops an absolute minimum when $b^2 > 3 \kappa / 2$.\label{fig:pot_min}}
\end{figure}

While developing this work, the authors of \cite{oikonomou} built an inflationary model in which the effective potential of the inflaton field has a Woods-Saxon form (i.e. a logistic function); here instead, using the ISB procedure \cite{chervon}, we start with a logistic {\it generating function} and from it we derive the potential and the Hubble parameter. We therefore define the function
\begin{equation}
	F \left( \phi \right) = \frac{A}{ 1 + \lebm },
\end{equation}
where we assume $A$ and $\lambda$ are positive constants, while $b$ is a nonzero constant. Notice that we can rewrite this as
\begin{equation}
	F \left( \phi \right) = \frac{ A e^{b\phi} }{ \lambda + e^{b\phi} }.
\end{equation}
It is also useful to notice that the generating function $F$ and its derivative $F^\prime$ satisfy $F^\prime = \frac{b}{A} F \left( A - F \right)$. \newline
\indent The potential $V$, the Hubble parameter $H$ and the inflaton field $\phi$ are related through \cite{chervon}:
\begin{eqnarray}
	H \left( \phi \right) & = & \sqrt{ \frac{\kappa}{3} } \left( F \left( \phi \right) + c \right) , \\
	V \left( \phi \right) & = & - \frac{2}{3\kappa} \left( F^\prime \left( \phi \right) \right)^2 + \left( F \left( \phi \right) + c \right)^2, \\
	H^\prime \left( \phi \right) & = & - \frac{\kappa}{2} \dot{\phi} ,
\end{eqnarray}
where $\kappa \equiv 1/M_P^2$ and the constant $c$ can in principle take any value. To simplify our analysis we will take $c=0$ \footnote{Taking $c \neq 0$ opens the possibility for the potential to develop a false vacuum depending on the values of the other parameters of the model.} so we simply have
\begin{equation}
	V \left( \phi \right) = - \frac{2A^2b^2\lambda^2}{3\kappa} \left[ \frac{ \ebm }{ \left( 1 + \lebm \right)^2 } \right]^2 + \frac{A^2}{ \left( 1 + \lebm \right)^2 }.
	\label{eq:potential_c0}
\end{equation}
Notice that this potential is quite different from a simple logistic function, and in fact opens the possibility for different shapes depending on the value of $b$. Thus, setting the constant $c=0$, the potential offers two distinct qualitative scenarios: it is easy to prove that for $b^2 \leq 3 \kappa / 2$ the potential has a well-known sigmoidal shape, as shown in figure \ref{fig:pot_sig}; on the other hand, if $b^2 > 3 \kappa / 2$, the potential develops a small curved well with a global minimum at around its center. This scenario can be visualized in figure \ref{fig:pot_min}.

\indent Also, notice that if we write the constant $\lambda$ as $\lambda \equiv e^{b\phi_0}$, we can see that its effect on the potential is to make the shift $\phi \rightarrow \phi - \phi_0$, thus amounting solely to a horizontal translation of the plot of $V$, but no change in the actual dynamics of the inflaton. Given this, we expect $\lambda$ not to play a determinant role in our results for inflationary and reheating observables; we will see that indeed this is the case throughout the course of this paper.

Notice that in the expression for our potential, the change $\phi \rightarrow -\phi$ is completely equivalent to changing $b \rightarrow -b$, and amounts to reflecting the plot of $V \left( \phi \right)$ around the vertical axis. Due to this symmetry we will focus only on positive values for $b$.

\indent We can write the Hubble parameter and the time derivative $\dot\phi$ as functions of $\phi$ \cite{chervon}:
\begin{eqnarray}
	H \left( \phi \right) & = & \sqrt{ \frac{\kappa}{3} } \frac{A}{ 1 + \lebm }, \\
	\dot{\phi} \left( \phi \right) & = & - \sqrt{ \frac{4}{3\kappa} } A b \lambda \frac{\ebm}{ \left( 1 + \lebm \right)^2 } .
\end{eqnarray}
These two expressions allow us to easily calculate the number of e-folds $N$ in terms of $\phi$ as
\begin{equation}
	N \equiv \int_{t_i}^{t_f} H dt = \int_{\phi}^{\phi_e} \frac{H}{\dot\phi} d\phi ,
\end{equation}
where $\phi_e$ denotes the value of the inflaton at the end of inflation. Substituting, we get
\begin{equation}
	N \left( \phi \right) = \frac{\kappa}{2b} \left. \left( \frac{1}{b\lambda} e^{b\phi} + \phi \right) \right|_{\phi_e}^{\phi}.
\end{equation}
Obviously the exact same answer is obtained if we were to calculate it as $N = \sqrt{\kappa} \int_{\phi_e}^\phi \frac{d\phi}{\sqrt{2\epsilon}}$. If we define
\begin{equation}
	\Phi \equiv \frac{2bN}{\kappa} + \frac{e^{b\phi_e}}{b\lambda} + \phi_e,
	\label{eq:Phi_def}
\end{equation}
then we can solve for $\phi$ simply as
\begin{equation}
	\phi = \Phi - \frac{1}{b} W \left( \frac{e^{b \Phi}}{\lambda} \right),
	\label{eq:phi_Phi}
\end{equation}
where $W$ is the Lambert function.

\indent The value of $\phi$ at the end of inflation can be calculated analytically directly from the condition $p < - \frac{1}{3} \rho$ and gives us:
\begin{equation}
	\phi_e = \frac{1}{b} \ln \lambda \left( \sqrt{\frac{2}{\kappa}} b - 1 \right).
	\label{eq:ph_e}
\end{equation}
In calculating this, we have assumed $b > 0$. Obviously, the same result is obtained by solving $\epsilon = 1$ for $\phi$ (see below). Notice that in order for inflation to end we must have $b > \sqrt{\frac{\kappa}{2}}$. It is very useful to define the dimensionless parameter $q$ through $b \equiv \sqrt{\frac{\kappa}{2}} \left( q + 1 \right)$. Thus, to have a well-defined value for $\phi_e$ we must have $q > 0$. Finally, using the property $W(x) e^{W(x)} = x$ it is straightforward to obtain the useful expression
\begin{equation}
	e^{b \phi} = \lambda W \left( q e^q e^{(q+1)^2N} \right).
	\label{eq:ephW}
\end{equation}

\section{Inflationary observables}

Slow roll indices are usually calculated using quotients of $V$ and its derivatives, giving a very good approximation as long as the field is slowly rolling, although that may be violated near the end of inflation. In our approach, it is actually possible (and easier) to work with the {\it exact} slow roll parameters, given by
\begin{eqnarray}
	\epsilon & = & - \frac{ \dot{H} }{ H^2 } , \\
	\eta & = & \frac{ \dot\epsilon }{ \epsilon H } .
\end{eqnarray}
Here, $\epsilon$ measures the rate of growth of the Hubble parameter and $\eta$ measures the rate of growth of $\epsilon$ itself. A straightforward calculation gives us
\begin{eqnarray}
	\epsilon & = & \frac{ 2 b^2 \lambda^2 }{ \kappa } \frac{1}{ \left( \lambda + e^{b\phi} \right)^2 } , \\
	\eta & = & \frac{ 4 b^2 \lambda }{ \kappa } \frac{ e^{b\phi} }{ \left( \lambda + e^{b\phi} \right)^2 } .
\end{eqnarray}
Plugging eq. \ref{eq:ephW} into these expressions gives us the slow roll parameters in terms of the number of e-folds.

\indent Given that we do not use slow roll approximation, the spectral index, tensor-to-scalar ratio and spectrum of scalar perturbations are given respectively as \cite{baumann} $n_s = 1 - 2 \epsilon_\ast - \eta_\ast$, $r = 16 \epsilon_\ast$, and $A_s = \frac{1}{8\pi^2} \frac{H_\ast^2}{\epsilon_\ast}$.
In our model these turn out to be
\begin{eqnarray}
	n_s & = & 1 - \frac{4 b^2 \lambda}{ \kappa \left( \lambda + e^{b\phi} \right) }, \\
	r & = & \frac{ 32 b^2 \lambda^2 }{ \kappa \left( \lambda + e^{b\phi} \right)^2 }, \\
	A_s & = & \frac{ A^2 \kappa^3 }{ 48 \pi^2 b^2 \lambda^2 } e^{2b\phi} ,
\end{eqnarray}
where the quantities on the right hand sides are to be evaluated at horizon exit. Using eq. \ref{eq:ephW} we can rewrite these as
\begin{eqnarray}
	n_s & = & 1 - \frac{ 2 \left( q + 1 \right)^2 }{ 1 + W \left( q e^q e^{ \left( q + 1 \right)^2 N_\ast } \right) }, \label{eq:ns_N} \\
	r & = & \frac{ 16 \left( q + 1 \right)^2 }{ \left[ 1 + W \left( q e^q e^{ \left( q + 1 \right)^2 N_\ast } \right) \right]^2 }, \\
	A_s & = & \frac{ A^2 \kappa^2 }{ 24 \pi^2 } \frac{ W^2 \left( q e^q e^{ \left( q + 1 \right)^2 N_\ast } \right) }{ \left( q + 1 \right)^2 }. \label{eq:As_A}
\end{eqnarray}

\begin{figure}[h!]
	\centering
	\includegraphics[scale=0.35]{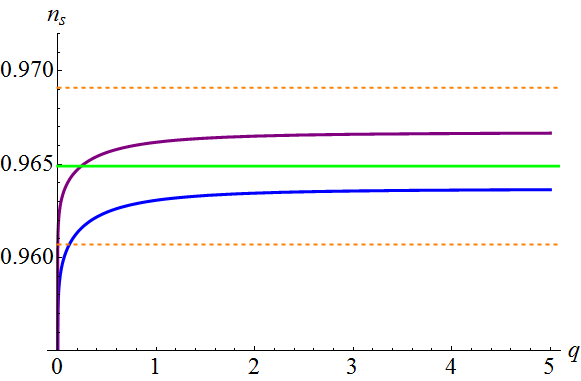}
	\caption{The spectral index. The green horizontal line corresponds to Planck's central value of $n_s = 0.9649$ while the orange dashed lines give the limits of 68\% confidence level interval. \label{ns}}
\end{figure}

\begin{figure}[h!]
	\centering
	\includegraphics[scale=0.30]{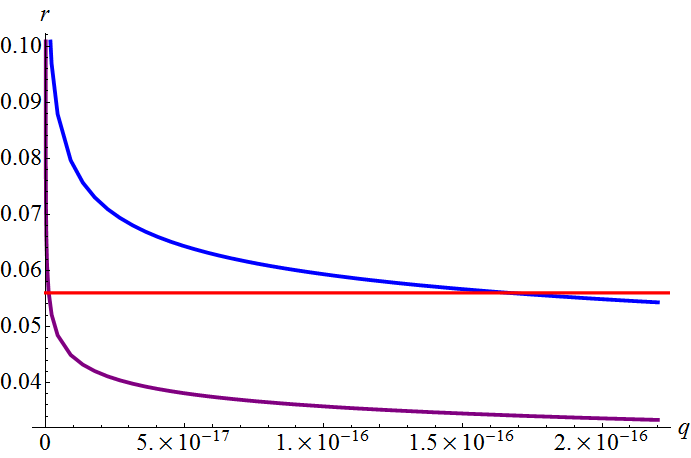}
	\caption{The tensor to scalar ratio. The red horizontal line gives the upper bound of $r < 0.056$. We see that our model satisfies this bound quite comfortably even for very small values of the parameter $q$, of the order $\sim 10^{-16}$.\label{r}}
\end{figure}

Notice that for the spectral index and the tensor-to-scalar ratio, the only parameter that plays a role is $q$ (or equivalently, $b$), while $\lambda$ completely disappears from the expressions; on the other hand, from eq. \ref{eq:As_A} we can get a bound on our parameter $A$ for given values of $N_\ast$, $q$ and using the central value of Planck collaboration's measurement of $A_s$. Plots of $n_s$ and $r$ can be seen in Fig. \ref{ns} and Fig. \ref{r} respectively. In both of these plots, the blue and purple curves correspond to $N_\ast = 55$ and $N_\ast = 60$ e-folds, respectively.

\section{Reheating}
Although reheating is still a mysterious epoch in the evolution of the universe, it is an almost essential period in the history of the universe in order for it to contain the kind of matter that we have today. As remarked in \cite{cook}, however, the reheating epoch of certain inflationary models can be characterized by the number of e-folds $N_R$ between the end of inflation and the start of the radiation-dominated era, the temperature $T_R$ at which thermalization between the inflaton and its decay products occur, and the equation of state $ \omega_R = p / \rho $ during reheating. \newline
\indent In \cite{cook} and \cite{munoz}, generic expressions for $N_R$ and $T_R$ are derived assuming an equation of state with constant $\omega_R$. In particular, assuming conservation of entropy between the reheating era and today, we have \cite{cook}
\begin{equation}
	N_R = \frac{4}{1-3\omega_R} \left[ - \frac{1}{4} \ln \left( \frac{45}{\pi^2g_R} \right) - \frac{1}{3} \ln \left( \frac{11g_R}{43} \right) - \ln \left( \frac{k}{a_0T_0} \right) - \ln \left( \frac{V_e^{1/4}}{H_\ast} \right) - N_\ast \right],
\end{equation}
for $\omega_R \neq 1/3$, where $g_R$ is the quantity of relativistic species at the end of the reheating phase, $k$ is a specific pivot scale, $a_0$ and $T_0$ are the scale factor and temperature at the present day respectively, and $V_e$ is the value of our potential evaluated at the end of inflation. Taking Planck's pivot scale of $k = 0.05$ Mpc$^{-1}$ and using the estimated value $g_R \approx 100$, this simplifies to \cite{cook}:
\begin{equation}
	N_R \approx \frac{4}{1-3\omega_R} \left[ 61.6 + \ln \left( \frac{H_\ast}{V_e^{1/4}} \right) - N_\ast \right],
	\label{eq:NR}
\end{equation}
where only the last two terms depend on our specific model. \newline
\indent One can invert eq. \ref{eq:ns_N} to express $N_\ast$ in terms of $n_s$:
\begin{equation}
	N_\ast = \frac{2}{1-n_s} - \frac{1}{q+1} + \frac{1}{(q+1)^2} \ln \left( \frac{2(q+1)^2}{q(1-n_s)} - \frac{1}{q} \right).
	\label{eq:N_ns}
\end{equation}
On the other hand, we can use eq. \ref{eq:As_A} to solve for $A$ in terms of $A_s$ and then obtain $H_\ast$ and $V_e$ by simply evaluating the Hubble parameter and the potential at $\phi_\ast$ and $\phi_e$ respectively:
\begin{eqnarray}
	H_\ast & = & \pi \sqrt{\frac{8A_s}{\kappa}} \frac{q+1}{ 1 + W \left( q e^q e^{(q+1)^2N_\ast} \right) }, \\
	V_e & = & \frac{8\pi^2A_s}{\kappa^2} \frac{2q^2-2q-1}{W^2 \left( q e^q e^{(q+1)^2N_\ast} \right)}.
\end{eqnarray}
Now we can plug eq. \ref{eq:N_ns} into these last two equations and substitute them into eq. \ref{eq:NR} to obtain $N_R$ in terms of $A_s$, $n_s$ and $q$; we do not write this expression explicitly because it is very long. \newline
\indent Finally, we can use the expression given in \cite{cook} for the reheating temperature:
\begin{equation}
	T_R = \left( \frac{43}{11g_R} \right)^{1/3} \frac{a_0}{k} T_0 H_\ast e^{-N_\ast} e^{-N_R},
\end{equation}
where we have already calculated all of the quantities on the right hand side.

\begin{figure}
	\centering
	\includegraphics[scale=0.3]{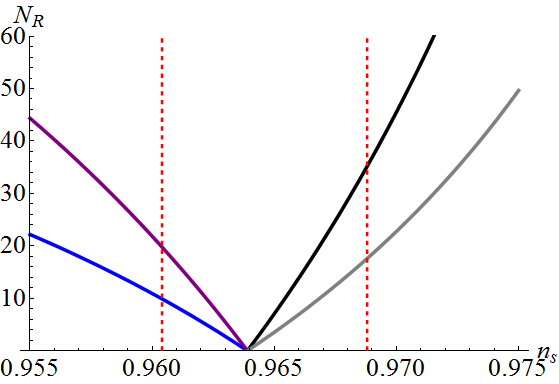}
	\caption{Our model predicts a relatively short duration of reheating, particularly compared to the results obtained in \cite{oikonomou}. Instantaneous reheating is achieved for a value of $n_s$ pretty close to Planck's central value of $0.9649$.\label{fig:nr}}
\end{figure}

\begin{figure}
	\centering
	\includegraphics[scale=0.3]{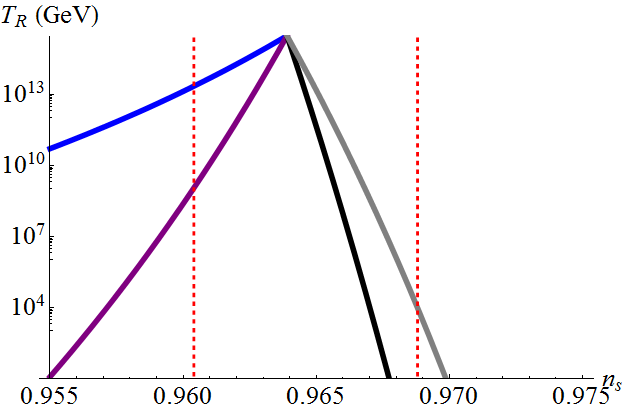}
	\caption{As seen in the previous graph for $N_R$, instantaneous reheating is obtained for a value of $n_s$ almost at the center of Planck's 68\% confidence level interval. Due to this fact, the predicted reheating temperatures are in general high. However, one can still have a wide range of possible values for $T_R$, especially if the equation of state assumed has $\omega_R > 0$.\label{fig:tr}}
\end{figure}

\indent We find that the value of the parameter $q$ does not significantly affect the $N_R$ and $T_R$ plots so we simply take $q=2$ for both figures presented in Fig. \ref{fig:nr} and Fig. \ref{fig:tr}. Also, in those graphs, the blue, purple, black and gray lines correspond to $\omega_R = -\frac{1}{3}$, $0$, $\frac{2}{3}$, and $1$ respectively. The dotted vertical lines correspond to the limits of the interval of 68\% confidence given by Planck \cite{Planck}.

\indent We feel that it is important to notice that the reheating temperature plots presented in \cite{oikonomou} are done by varying their parameter $V_0$ (which is the analogous of our $A^2$ in the sense that they control the ``amplitude'' of the potential) over a huge range of values, from $\sim 10^{2}$ to $\sim 10^{-60}$; but since in their model the scalar power spectrum is directly proportional to $V_0$, the variation of $V_0$ over such a large range would affect their prediction for $A_s$, which would be off by several orders of magnitude.

\section{Conclusion}
Using the ISB procedure, we have analytically calculated a new inflationary potential, from which we successfully obtain a spectral index of scalar perturbations, the tensor-to-scalar ratio and cosmological reheating that satisfy current experimental bounds. Given these virtues, we recognize that it can be seriously considered as a new model of cosmological inflation.

\end{document}